\documentclass[11pt,bm,aps,showpacs,nofootinbib,amsfonts,amssymb,preprintnumbers]{revtex4}

 \usepackage{here}

\usepackage[dvipdfmx]{graphicx}
\usepackage{amssymb,amsfonts,amsmath,bm,color,multirow,cases,empheq,hyperref}
\usepackage{mathrsfs}

\usepackage{enumerate}
\usepackage{ulem}
\usepackage{afterpage}

\hypersetup{%
 setpagesize=false,
 bookmarksnumbered=true,%
 bookmarksopen=true,%
 colorlinks=true,%
 linkcolor=blue,%
 citecolor=red}

\begin{document}




\title{
Stable bound orbits in microstate geometries
}

\author{Shinya Tomizawa}
\email{tomizawa@toyota-ti.ac.jp}
\author{Ryotaku Suzuki}
\email{sryotaku@toyota-ti.ac.jp}
\affiliation{Mathematical Physics Laboratory, Toyota Technological Institute\\
Hisakata 2-12-1, Nagoya 468-8511, Japan}
\date{\today}

\preprint{TTI-MATHPHYS-12}




\begin{abstract} 
 We show the existence of  stable bound orbits for the massive and massless particles moving in the simplest microstate geometry spacetime in the bosonic sector of the five-dimensional minimal supergravity. 
 In our analysis, reducing the motion of particles to a two-dimensional potential problem, we numerically investigate whether the potential has a negative local minimum. 
 \end{abstract}

\date{\today}
\maketitle



\section{Introduction}

The microstate geometries~
\cite{Lunin:2001jy,Maldacena:2000dr,Balasubramanian:2000rt,Lunin:2002iz,Lunin:2004uu,Giusto:2004id,Giusto:2004ip,Giusto:2004kj,Bena:2005va,Gibbons:2013tqa,Saxena:2005uk,Skenderis:2008qn,Balasubramanian:2008da,Chowdhury:2010ct} are smooth horizonless solutions with the same asymptotic structure as a black hole or a black ring. 
So far, these solutions have been thought of as one of ways to resolve  the problem of black hole information loss. 
The idea to describe black hole microstates by horizonless geometries first originated from the works on fuzzballs of Mathur~\cite{Mathur:2005zp,Mathur:2005ai,Mathur:2008nj}.  
The existence itself of such solutions should be surprising because in the earlier works \cite{Einstein1941,Einstein1943,Breitenlohner:1987dg,Cater1986}, it is a well-known fact that smooth soliton solutions in four dimensions are completely excluded. 
However, in five dimensions, at least, in five dimensional supergravity, the no-go theorem does not hold  because the spacetime admits the spatial cross sections with non-trivial second homology and the Chern-Simons interactions.

\medskip
There are many ways to probe physical aspects of such microstate geometries. 
The simplest and most interesting way to probe the microstate geometries may be studying geodesic motion of massive and massless particles in such spacetimes.  
In particular it is an interesting issue what is significantly different between the motion of particles (e.g. the existence/nonexistence of stable bound orbits) in microstate geometries and that  in black hole spacetimes. 
Many researchers have so far studied particle motions  in  black hole spacetimes.
For example, it is well-known that in a four-dimensional Schwarzschild background, stable bound orbits exist for massive particles and do not  exist for massless particles, whereas 
in a five-dimensional Schwarzschild background, they do not exist for both massive and massless  particles.  
Moreover, stable bound orbits for rotating spherical black holes~\cite{Wilkins:1972rs,Frolov:2003en,Diemer:2014lba,Igata:2014xca}, black holes with non-trivial topologies~\cite{Igata:2010ye,Igata:2010cd,Igata:2013be,Tomizawa:2019egx,Tomizawa:2020mvw,Nakashi:2019mvs,Igata:2020vlx} and Kaluza-Klein black holes~\cite{Igata:2021wwj,Tomizawa:2021vaa} were also investigated. 
On the other hand, 
in supersymmetric microstate geometries~\cite{Eperon:2016cdd,Eperon:2017bwq}, massless particles with zero energy are stably trapped on an evanescent ergosurface, which are defined as timelike hypersurfaces such that a stationary Killing vector field becomes null there and timelike everywhere except there.

\medskip
As shown in ref.~\cite{Tomizawa:2021qli}, the angular momenta of the microstate geometries with a small number of centers (at least, five centers) have lower bounds, which are slightly smaller than those of the maximally spinning BMPV black holes for asymptotically flat, stationary solutions with bi-axisymmetry and reflection symmetry in the five-dimensional ungauged minimal supergravity.
This means that there exists a certain  parameter region such as the microstate geometries with a small number of the centers have the same angular momenta as the BMPV black holes.  
It is interesting to compare particle motion in the microstate geometries within such a parameter region with one in the black hole spacetime. 
For instance, if the geodesic motions of particles in the two spacetimes significantly differ, the microstate geometries may be unable to be regarded as an alternative to the black hole.  
The main purpose of this paper is to study stable bound orbits in the microstate geometries, 
more precisely, to investigate numerically whether stable bound orbits of particles can exist in the microstate geometries with the same mass and angular momenta as the BMPV black holes.

\medskip

In our analysis, focusing on for asymptotically flat, stationary BPS solutions with bi-axisymmetry and reflection symmetry in the five-dimensional ungauged minimal supergravity, we regard motion of particles as a two-dimensional potential problem.  
As discussed for several BPS black holes in the same theory~\cite{Tomizawa:2019egx,Igata:2020vlx,Tomizawa:2020mvw,Igata:2021wwj,Tomizawa:2021vaa}, one can replace a problem of whether there exist stable bound orbits for particles with a simple problem of whether the two-dimensional effective potential has a negative local minimum. 
First, we numerically show the existence of stable bound orbits for massive particles for the microstate geometries with three Gibbons-Hawking centers. 
Next, we numerically show that there can be stable bound orbits of massive and massless particles for the microstate geometries with five Gibbons-Hawking centers which has the same mass and angular momenta as the BMPV black holes. 
Moreover, we also study the five center solutions whose angular momenta are larger than the BMPV black holes.   

\medskip
The rest of the paper is organized as follows: 
In the following Sec.~\ref{sec:msg}, we briefly review the microstate geometries in the five-dimensional minimal supergravity. 
In Sec.~\ref{sec:formalism}, we provide our formalism to show the existence of stable bound orbits. 
In Sec.~\ref{sec:SBO}, using the formalism, we discuss whether there are stable bound orbits for massive and massless particles.
In Sec.~\ref{sec:summary}, we summarize our results and discuss possible generalizations of our analysis.



\section{Microstate geometries}\label{sec:msg}

\subsection{Solutions}

We review the microstate geometries in the five-dimensional minimal ungauged supergravity~(the five-dimensional Einstein-Maxwell theory with a Chern-Simons term)~\cite{Gauntlett:2002nw,Tomizawa:2021qli}.
 The metric and the gauge potential $1$-form of the Maxwell field can be written as
\begin{eqnarray}
\label{metric}
ds^2&=&-f^2(dt+\omega)^2+f^{-1}ds_{M}^2,\label{eq:solution_metric}\\
A&=&\frac{\sqrt 3}{2} \left[f(d t+\omega)-\frac KH(d \psi+\chi)-\xi \right], \label{eq:solution_1form}
\end{eqnarray}
where $ds^2_M$ is the metric of the Gibbons-Hawking space~\cite{Gibbons:1979zt}, which is written as
\begin{eqnarray}
ds^2_M&=&H^{-1}(d\psi+\chi)^2+Hds^2_{{\mathbb E}^3}, \label{eq:GH}\\
ds^2_{{\mathbb E}^3}&=&dx^2+dy^2+dz^2=dr^2+r^2(d\theta^2+\sin^2\theta d\phi^2),\\
\chi&=&\sum_{i=1}^nh_i \frac{r\cos\theta -z_i}{r_i}d\phi,\\
H&=&\sum_{i=1}^n\frac{h_i}{r_i}. \label{Hdef}
\end{eqnarray}
Here, $r_i\ (i=1,\ldots ,n)$ is the distance between ${\bm r}:=(x,y,z)$ and the $i$-th point source  ${\bm r}_i:=(0,0,z_i)$ of a harmonic function $H$ on ${\mathbb E}^3$ and is defined as
\begin{eqnarray}
r_i:=|{\bm r}-{\bm r_i}|=\sqrt{x^2+y^2+(z-z_i)^2}=\sqrt{r^2-2rz_i\cos\theta+z_i^2}.
\end{eqnarray}
The function $f^{-1}$ and the 1-forms $(\omega,\xi)$ are written as
\begin{eqnarray}
f^{-1}&=&H^{-1}K^2+L,\\
\omega&=& \left( H^{-2}K^3+\frac{3}{2} H^{-1}KL+M \right)(d\psi+\chi)\notag \\
&&+
\biggl[\sum_{i,j=1\ (i\not=j)}^n\left(h_i m_j+\frac{3}{2}k_il_j\right)\frac{r^2-(z_i+z_j)r\cos\theta+z_iz_j}{z_{ji}r_ir_j}\notag\\
&&-\sum_{i=1}^n\left( m_0h_i+\frac{3}{2}l_0k_i \right)\frac{r\cos\theta -z_i}{r_i}+c\biggr]d\phi,\\
\xi&=&-\sum_{i=1}^n k_i \frac{r\cos\theta -z_i}{r_i}d\phi,
\end{eqnarray}
where the functions $K$, $L$ and $M$ are harmonic functions on ${\mathbb E}^3$, 
\begin{eqnarray}
K=\sum_{i=1}^n\frac{k_i}{r_i},\ L=l_0+\sum_{i=1}^n\frac{l_i}{r_i}, \ 
M=m_0+\sum_{i=1}^n\frac{m_i}{r_i}.
\end{eqnarray}
Since we assume that all point sources lie on the $z$-axis, the solutions have three  commuting Killing vectors 
 $\partial/\partial t\ ,\partial/\partial \psi$ and $\partial/\partial\phi$, and 
 the coordinate has the ranges $-\infty<t<\infty$, $r>0$, $0\le\psi <4\pi$, $0\le \phi<2\pi$ and $0\le \theta\le \pi$.

\medskip



As discussed in Ref.~\cite{Tomizawa:2016kjh,Tomizawa:2021qli}, asymptotic flatness requires the parameters to be subject to
\begin{eqnarray}
\sum_{i=1}^n h_i&=& 1,\\
l_0&=&1,\\
c&=&\sum_{i,j=1(i\not=j)}^n\frac{h_im_j+\frac{3}{2}k_il_j}{z_{ij}},\\
m_0&=&-\frac{3}{2}\sum_{i=1}^nk_i.
\end{eqnarray}
From the requirements of regularity at ${\bm r}={\bm r}_i\ (i=1,\ldots,n)$, the parameters $(k_{i\ge 1},l_{i\ge 1},m_{i\ge 1})$ must satisfy 
\begin{eqnarray}
l_i&=&-\frac{k_i^2}{h_i},\\
m_i&=&\frac{k_i^3}{2h_i^2}.
\end{eqnarray}
Moreover, to ensure Lorenzian signature of the metric around the points ${\bm r}={\bm r}_i\ (i=1,\ldots,n)$, the inequalities 
\begin{eqnarray}
h_i^{-1}c_{1(i)} &:=&h_i+\sum_{j=1(j\not=i)}^n \frac{2k_ik_j+l_ih_j+h_il_j}{|z_{ij}|} >0,\label{eq:c1}
\label{eq:c1}
\end{eqnarray}
must be satisfied, and from the absence of closed timelike curves around the points, they must also satisfy
\begin{eqnarray}
h_ic_{2(i)}&:=&h_im_0+\frac{3}{2}k_i+\sum_{j=1(j\not=i)}^n\frac{h_im_j-m_ih_j-\frac{3}{2}(l_ik_j-k_il_j)}{|z_{ij}|}=0.
\label{eq:c2}
\end{eqnarray}
In addition, the absence of orbifold singularities at ${\bm r}={\bm r}_i\ (i=1,\ldots,n)$ demands
\begin{eqnarray}
h_i=\pm 1\ (i=1,\ldots,n) .
\end{eqnarray}

\subsection{Useful coordinates}
In the work of the geodesic motion of particles, it is more convenient to use the coordinates $(\rho,\phi_1,\phi_2)$ defined by
\begin{eqnarray}
x=\rho \cos(\phi_1-\phi_2),\quad y=\rho \sin(\phi_1-\phi_2),\quad \psi=\phi_1+\phi_2,\quad \phi=\phi_1-\phi_2,
\end{eqnarray}
where $(\phi_1,\phi_2)$ are the coordinates with $2\pi$ periodicity.

\subsection{Three-center solutions}
 The solutions with three centers ($n=3$) and $(h_1,h_2,h_3)=(1,-1,1)$ describe the simplest asymptotically flat, stationary and bi-axisymmetric microstate geometries, which have the four parameters $(k_1,k_3,z_1,z_3)$, where we have set $k_2=0$ and $z_2=0$. 
Moreover, under the assumption of the reflection symmetry 
\begin{eqnarray}
z_3=-z_1=:a\ (>0),\qquad k_3=k_1, \label{eq:reflection_n3}
\end{eqnarray}
the bubble equations~(\ref{eq:c2}) are simply written as 
\begin{eqnarray}
c_{2(1)}=-\frac{1}{2}c_{2(2)}=c_{2(3)}=\frac{k_1[k_1^2-3a]}{2a}=0, \label{eq:c2_n3}
\end{eqnarray}
which lead to
\begin{eqnarray}
k_1=0,\quad a=\frac{k_1^2}{3}.   \label{eq:c2l}
\end{eqnarray}
Only the latter case can satisfies the inequalities~(\ref{eq:c1}), where $h_ic_{1(i)}\ (i=1,2,3)$ are computed as
\begin{eqnarray}
h_1c_{1(1)}=h_3c_{1(3)}=4,\quad h_2c_{1(2)}=5.
\end{eqnarray}
Therefore, for arbitrary nonzero $k_1$, this describes regular and causal solutions of asymptotically flat, stationary microstate geometries with the bi-axisymmetry and reflection symmetry. 
The solutions were previously analyzed in Ref.~\cite{Gibbons:2013tqa}. 

\begin{figure}[H]
\centering
\includegraphics[width=8cm]{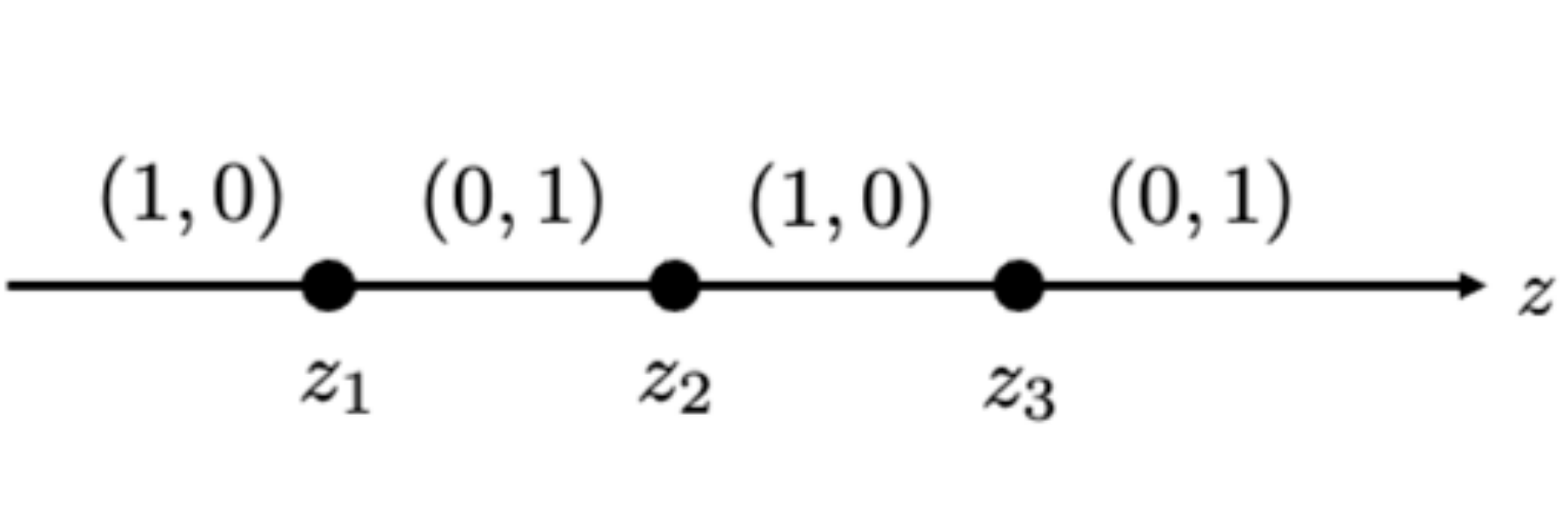}
\caption{Rod structure for the microstate geometries with three centers and $(h_1,h_2,h_3)=(1,-1,1)$.}
\label{fig:rod_n3}
\end{figure}

\medskip
The $z$-axis of ${\mathbb E}^3$ in the Gibbons-Hawking space consists of the four intervals: 
$I_-=\{(x,y,z)|x=y=0,  z<z_1\}$, $I_i=\{(x,y,z)|x=y=0,z_i<z<z_{i+1}\}\ (i=1,2)$ and $I_+=\{(x,y,z)|x=y=0,z>z_3\}$. 
The rod structure of the three-center microstate geometries is displayed in Fig.~\ref{fig:rod_n3}.

\medskip
 Under the symmetric conditions~(\ref{eq:reflection_n3}) and gauge conditions $k_2=0$, $z_2=0$, the ADM mass, two ADM angular momenta and  the magnetic fluxes on $I_i\ (i=1,2)$ are written as, respectively, 
\begin{eqnarray}
M&=&\frac{\sqrt{3}}{2}Q=6\pi k_1^2,\label{eq:mass_n3}\\
J_\psi&=&3\pi k_1^3,\label{eq:jpsi_n3}\\
J_\phi&=&0,\label{eq:jphi_n3}\\
q[I_1]&=&-q[I_2]=\frac{\sqrt{3}}{2}k_1.
\end{eqnarray}
As discussed in ref.~\cite{Tomizawa:2021qli}, 
the squared angular momentum normalized by the mass is written as
\begin{eqnarray}
j^2=\frac{9}{8}.
\end{eqnarray}
This is  larger than the BMPV black holes which have the range of $0\le j^2 <1$.

\subsection{Five-center solutions}
The stationary, bi-axisymmetric  microstate geometries with five centers ($n=5$) and  $(h_1,h_2,h_3,h_4,h_5)=(1,-1,1,-1,1)$ have the four parameters $(k_1,k_2,z_1,z_2)$ under the reflection-symmetric conditions
\begin{eqnarray}
k_5=k_1,\quad k_4=k_2,\quad z_5=-z_1=:a+b,\quad z_4=-z_2=:b
\end{eqnarray}
and the gauge conditions $k_3=0$, $z_3=0$. 
Thus, the conditions~(\ref{eq:c2}) are simplified  as
\begin{eqnarray}
&&2h_1c_{2(1)}=2h_5c_{2(5)}=-3(k_1+2k_2)-\frac{k_1^3}{a+b}+\frac{(k_1+k_2)^3}{a}+\frac{(k_1+k_2)^3}{a+2b}=0, \label{eq:c21_n5}\\
&&2h_2c_{2(2)}=2h_4c_{2(4)}=3(2k_1+3k_2)-\frac{k_2^3}{b}-\frac{(k_1+k_2)^3}{a}-\frac{(k_1+k_2)^3}{a+2b}=0, \label{eq:c22_n5}\\
&&h_3c_{2(3)}=-3(k_1+k_2)+\frac{k_1^3}{a+b}+\frac{k_2^3}{b}=0,\label{eq:c23_n5}
\end{eqnarray}
where we note that Eqs.~(\ref{eq:c21_n5})-(\ref{eq:c23_n5}) are not independent due to the constraint equation $\sum_{i=1}^5 h_ic_{2(1)}=2h_1c_{2(1)} +2h_2c_{2(2)}+h_3c_{2(3)}=0$. 
Therefore, the solutions have only two independent parameters. 
If we regard $a$ and $b$ as the functions of $k_1$ and $k_2$ from Eqs.~(\ref{eq:c21_n5}), (\ref{eq:c23_n5}),  the solutions are a two-parameter family for $(k_1,k_2)$. 

\begin{figure}[H]
\centering
\includegraphics[width=8cm]{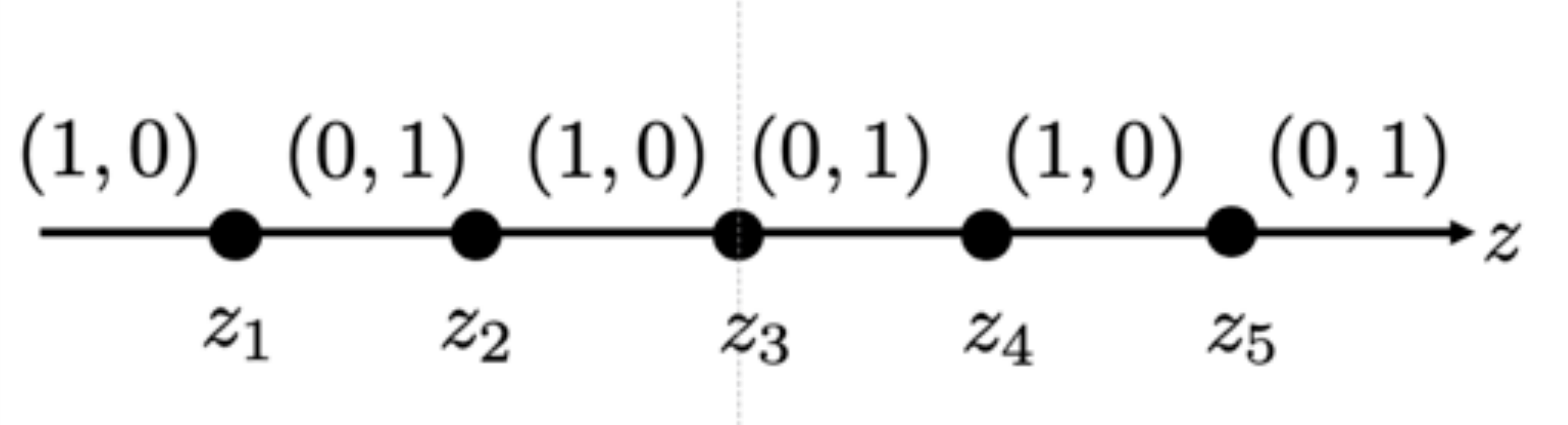}

\caption{Rod structure for the microstate geometries with five centers and $(h_1,h_2,h_3,h_4,h_5)=(1,-1,1,-1,1)$.}

\label{fig:rod_n5}
\end{figure}

Furthermore, the parameters $k_1$ and $k_2$ must satisfy the inequalities (\ref{eq:c1}), which are written as
\begin{eqnarray}
&&h_1c_{1(1)}=h_5c_{1(5)}=1-\frac{k_1^2}{a+b}+\frac{(k_1+k_2)^2}{a}+\frac{(k_1+k_2)^2}{a+2b}>0,\label{eq:c15}\\
&&h_2c_{1(2)}=h_4c_{1(4)}=-1+\frac{k_2^2}{b}+\frac{(k_1+k_2)^2}{a}+\frac{(k_1+k_2)^2}{a+2b}>0,\label{eq:c14}\\
&&h_3c_{1(3)}=1-\frac{2k_1^2}{a+b}+\frac{2k_2^2}{b}>0,\label{eq:c13_n5}
\end{eqnarray}
with the inequalities $a>0,\ b>0$.
In the below, we assume $k_1\not=0$ and $k_2\not=0$. 
As shown in~Ref.~\cite{Tomizawa:2021qli}, these inequalities are equivalent with 
\begin{eqnarray}
k_2/k_1<-1, \quad  -0.2063...<k_2/k_1<0, \quad  k_2/k_1>0. \label{eq:range}
\end{eqnarray}
The $z$-axis of ${\mathbb E}^3$ in the Gibbons-Hawking space consists of the six intervals: 
$I_-=\{(x,y,z)|x=y=0,  z<z_1\}$, $I_i=\{(x,y,z)|x=y=0,z_i<z<z_{i+1}\}\ (i=1,...,4)$ and $I_+=\{(x,y,z)|x=y=0,z>z_5\}$. 
The five-center microstate geometries have the rod structure displayed in Fig.~\ref{fig:rod_n5}.

\medskip
For the solutions, the ADM mass, two ADM angular momenta  the magnetic fluxes are computed as
\begin{eqnarray}
M&=&\frac{\sqrt{3}}{2}Q=6\pi( k_1^2+4k_1k_2+3k_2^2),\label{eq:mass_n5}\\
J_\psi&=&3\pi (k_1^3+6k_1^2k_2+10k_1k_2^2+5k_2^3),\label{eq:jpsi_n5}\\
J_\phi&=&0, \label{eq:jphi_n5}\\
q[I_1]&=&-q[I_4]=\frac{\sqrt{3}}{2}(k_1+k_2),\quad q[I_2]=-q[I_3]=-\frac{\sqrt{3}}{2}k_2.
\end{eqnarray}

\medskip
As discussed in ref.~\cite{Tomizawa:2021qli}, 
the squared angular momentum runs the range 
\begin{eqnarray}
j^2>0.841...\ . 
\end{eqnarray}
The bi-axisymmetric and reflectionally symmetric microstate geometries with five centers can have the  same angular momentum of the range $0.841...<j^2<1$ as the BMPV black holes.

\section{Our formalism} \label{sec:formalism}
To study stable bound orbits,  we regard the geodesic motion of massive and massless particles as a two-dimensional potential problem (see ref.\cite{Tomizawa:2019egx} about the detail). The Hamiltonian of a free particle with mass $m$ is written as
\begin{eqnarray}
\mathcal{H}=g^{\mu\nu}p_\mu p_\nu+m^2, \label{eq:Hamiltonian}
\end{eqnarray}
where $p_\mu$ is the momentum such that $(p_t,p_{\phi_1},p_{\phi_2})=(-E,L_{\phi_1},L_{\phi_2})$ are constants of motion.
Then, the Hamiltonian can be rewritten as
\begin{eqnarray}
\mathcal{H}=\frac{4f}{H\rho^2}(p_\rho^2+p_z^2)+E^2\left(U+\frac{m^2}{E^2}\right).\label{eq:Hamiltonian2}
\end{eqnarray}
The effective potential $U=U(\rho,z)$ is given by
\begin{eqnarray}
U&=&g^{tt}+g^{\phi_1\phi_1}l_{\phi_1}^2+g^{\phi_2\phi_2}l_{\phi_2}^2-2g^{t\phi_1} l_{\phi_1}-2g^{t\phi_2} l_{\phi_2}+2g^{\phi_1\phi_2}l_{\phi_1} l_{\phi_2}\\
&=&\frac{1}{4(K^2+HL)}\biggl[ -3K^2L^2+8K^3M+12HKLM-4HL^3+4H^2M^2 \notag\\
&&+ (4K^3+6HKL+4H^2M)(l_{\phi_1}+l_{\phi_2})+H^2(l_{\phi_1}+l_{\phi_2})^2 \biggr]\notag\\
&&+\frac{[-2\hat\omega_\phi +(l_{\phi_1}+l_{\phi_2})\chi_\phi+(l_{\phi_2}-l_{\phi_1})]^2}{4(K^2+HL)\rho^2},
\end{eqnarray}
where  two angular momenta $(L_{\phi_1},L_{\phi_2})$ are normalized by the energy $E$ as
 $l_{\phi_1}:=L_{\phi_1}/E$ and $l_{\phi_2}:=L_{\phi_2}/E$. 
Thus,  particles move on the two-dimensional space $(\rho,z)$ in the two-dimensional potential $U(\rho,z)$, satisfying the Hamiltonian constraint ${\mathcal H}=0$.
 The allowed regions of the motions for massive and massless particles correspond to $U\le -m^2/E^2$ and $U\le0$, respectively. 
If $U$ has a negative local minimum for given $(l_{\phi_1},l_{\phi_2})$, stable bound orbits exist for massive particles at the point or in the neighborhood of the point, and furthermore if the curve $U=0$ in the $(\rho,z)$-plane or the region surrounded with the $U=0$ curve and the $z$-axis  is closed, stable bound orbits exist even for massless particles.

\section{Stable bound orbits}\label{sec:SBO}

\subsection{Three-center solutions}
To see whether there exist stable bound orbits in the microstate geometries for $n=3$, let us focus on the simple case of $l_{\phi_2}=0$. 
For the three-center solutions, the $z$-axis is composed of the $4$ intervals, 
$I_-=\{(\rho,z)|\rho=0,z<z_1\}$, $I_1=\{(\rho,z)|\rho=0,z_1<z<z_{2}\}$, $I_2=\{(\rho,z)|\rho=0,z_2<z<z_{3}\}$ and $I_+=\{(\rho,z)|\rho=0,z>z_3\}$. 
As was previously shown in~\cite{Tomizawa:2019egx}, it should be noted that only particles with the angular momentum of $l_{\phi_2}=0$ can move on $I_1$ and $I_+$
(because $I_+$ and $I_1$ correspond to the fixed points of the Killing isometry $v_2:=\partial/\partial\phi_2$), and hence only particle with the angular momentum of $J:=p_\mu v_2^\mu=L_{\phi_2}=0\ (l_{\phi_2}=0)$ can move on the $z$-axis, whereas  (because $U$ diverges) on $I_-$ and $I_2$ the particles cannot move. 
It can be shown from the entirely similar discussion that on $I_-$ and $I_2$ only particles with the angular momenta of $l_{\phi_1}=0$ are allowed to move there.

\medskip

\begin{figure}[H]
 \begin{tabular}{ccc}
 \begin{minipage}[t]{0.3\hsize}
\includegraphics[width=5cm,height=5cm]{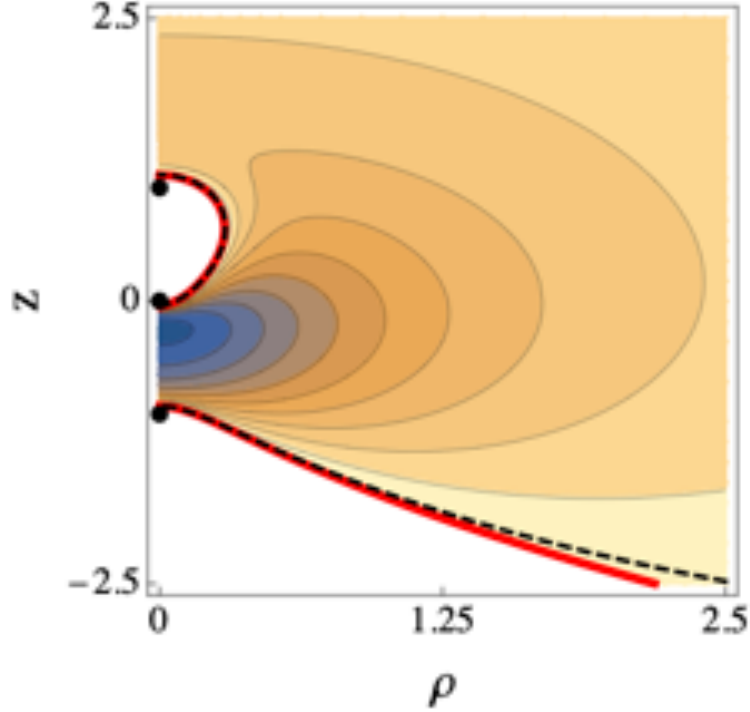}
 \end{minipage} & 
 
 \begin{minipage}[t]{0.3\hsize}
\includegraphics[width=5cm,height=5cm]{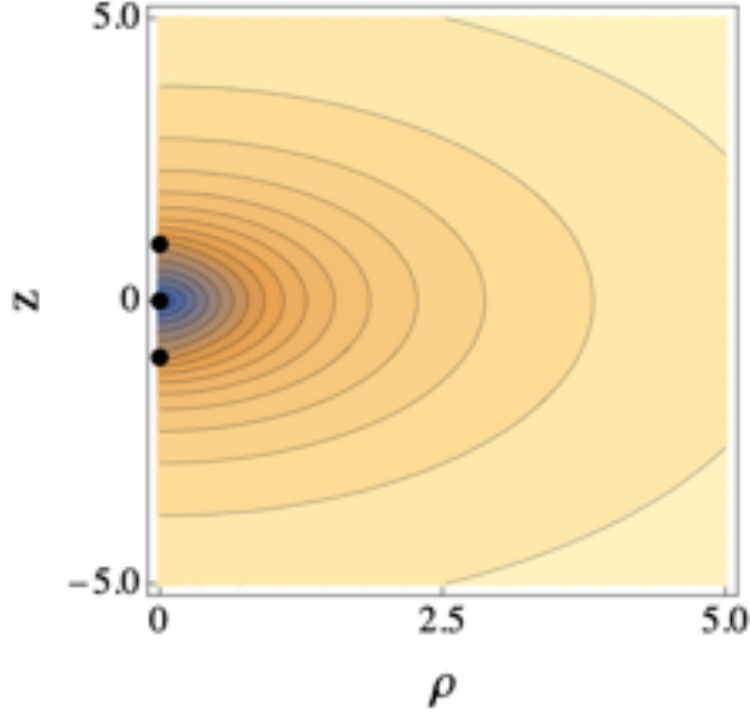}
 \end{minipage} &
 
 \begin{minipage}[t]{0.3\hsize}
 \includegraphics[width=5cm,height=5cm]{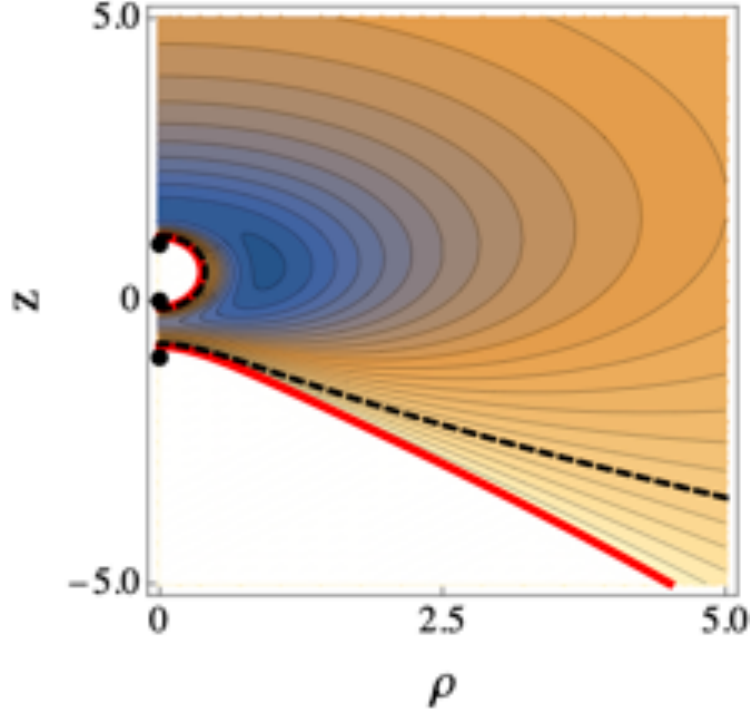}
 \end{minipage}\\
 
  \begin{minipage}[t]{0.3\hsize}
 \includegraphics[width=5cm,height=5cm]{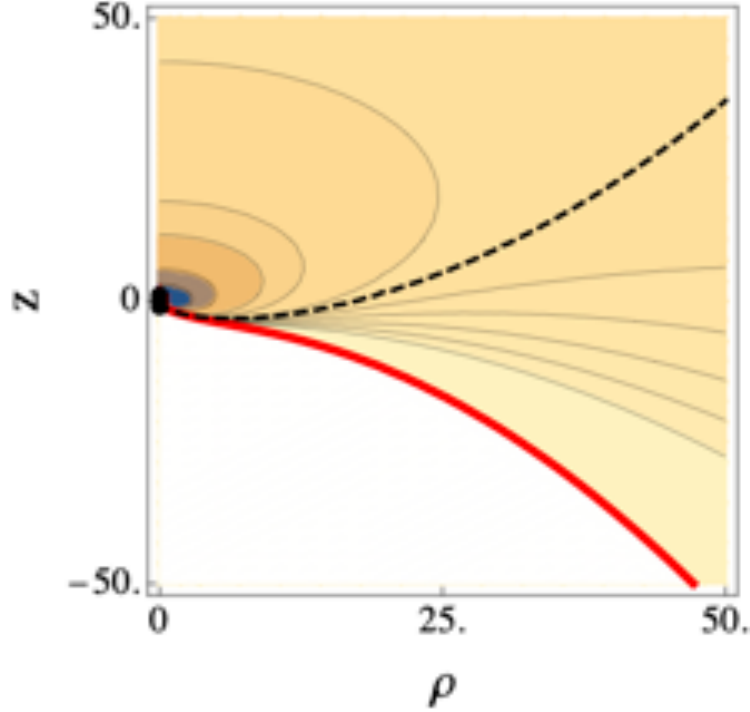}
 \end{minipage} & 
 
 \begin{minipage}[t]{0.3\hsize}
\includegraphics[width=5cm,height=5cm]{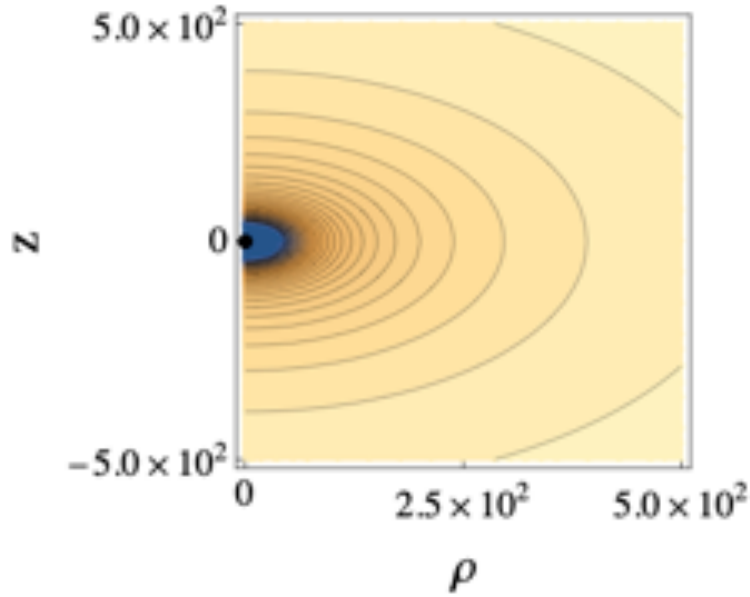}
 \end{minipage} &
 
 \begin{minipage}[t]{0.3\hsize}
 \includegraphics[width=5cm,height=5cm]{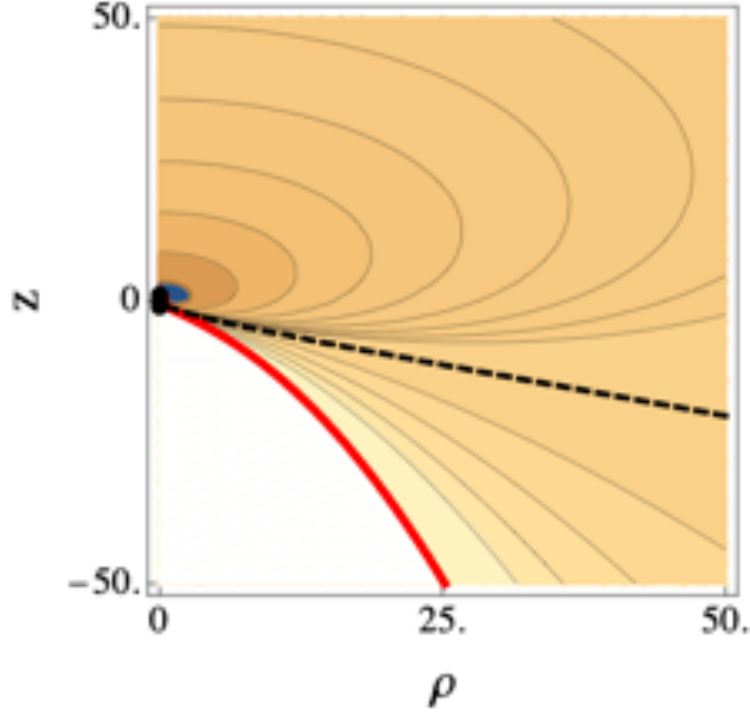}
 \end{minipage}\\
 
   \begin{minipage}[t]{0.3\hsize}
 \includegraphics[width=5cm,height=5cm]{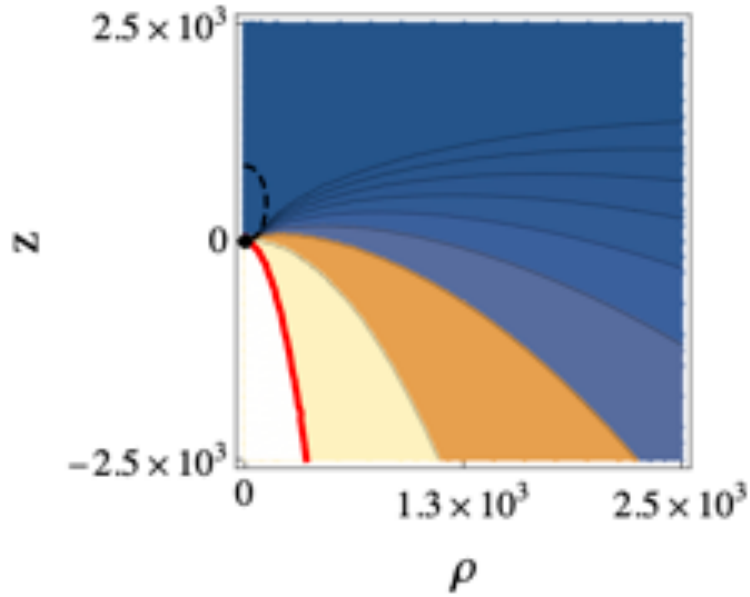}
 \end{minipage} & 
 
 \begin{minipage}[t]{0.3\hsize}
\includegraphics[width=5cm,height=5cm]{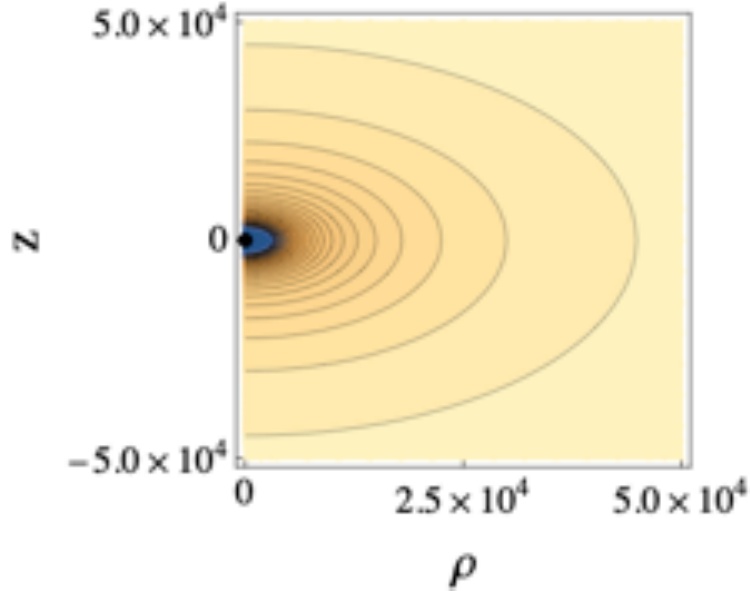}
 \end{minipage} &
 
 \begin{minipage}[t]{0.3\hsize}
 \includegraphics[width=5cm,height=5cm]{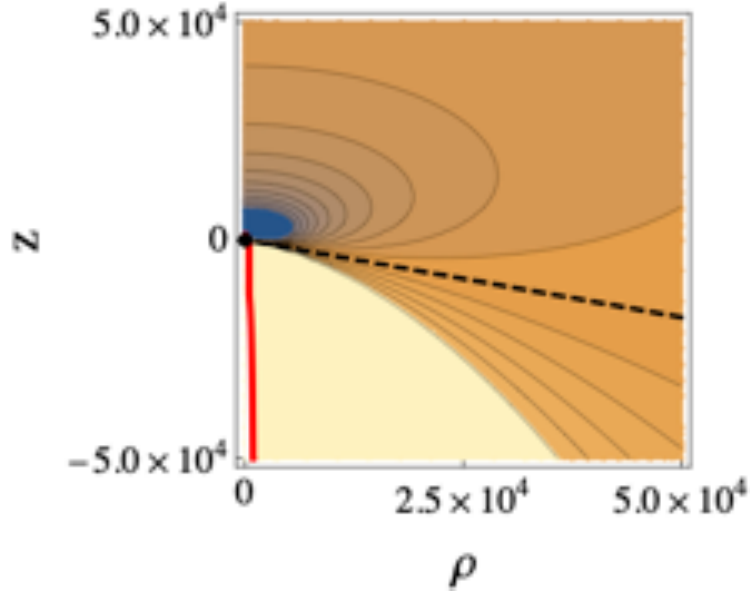}
 \end{minipage}

 \end{tabular}
\caption{The figures show the contours of the potential $U$ under the parameter setting $(k_1,k_2,k_3)=(\sqrt{3},0,\sqrt{3})$ and $(z_1,z_2,z_3)=(-1,0,1)$. 
The left, middle and right  figures correspond to the angular momenta $(l_{\phi_1},l_{\phi_2})=(-7,0), (0,0), (4,0)$,  respectively, and the upper, middle and lower figures are differs only in the scales of the vertical and horizontal axes.   
 The bold solid curves and the dashed curves denote $U=0$ and $U=-1$, respectively, and 
 the white regions denote the forbidden regions of $U>0$ where massive and massless particles cannot move. 
 The black circles correspond to the centers at $(\rho,z)=(0,z_i)\ (i=1,2,3)$.  } 
\label{fig:n3U}
\end{figure}

Figure \ref{fig:n3U} displays the typical contour plots of $U$ in the $(\rho,z)$-plane for the parameter-setting $(k_1,k_2,k_3)=(\sqrt{3},0,\sqrt{3})$, which corresponds to the solutions with $j^2=9/8$ and  $(z_1,z_2,z_3)=(-1,0,1)$, where 
it should be noted that the upper, middle and lower figures  differ only in the scales of the horizontal axis and vertical axis, and the left, middle and right figures correspond the angular momenta of particles $(l_{\phi_1},l_{\phi_2})=(-7,0), (0,0),(4,0)$, respectively.
In these figures, the red curves denote  the contours $U=0$, which we call $U=0$ curves. 
For each case of $(l_{\phi_1},l_{\phi_2})=(-7,0), (4,0)$, the $U=0$ curve intersects with  $I_1$ and $I_+$ on the $z$-axis so that it makes a closed region surrounded with the $z$-axis. 
Since $U>0\ (<0)$ inside (a little outside) the $U=0$ curve, there are not stable bound orbits  for massless particles.  
It can be seen from these figures that  in each case, $U$ has a negative local minimum outside the $U=0$ curve, i.e., 
there are stable bound orbits for massive particles. 
In particular, for $(l_{\phi_1},l_{\phi_2})=(-7,0)$,  the stable bound orbit at the local minimum is circular because $U$ has the local minimum on $I_1$, where massive particles move along $\partial/\partial \phi_1$. 
On the other hand, it can be shown from the three middle figures that  for $(l_{\phi_1},l_{\phi_2})=(0,0)$, $U$ does not have a $U=0$ curve and has a local minimum at the center $(0,0)$.  
Therefore,  stable bound orbits do not exist for massless particles but exist for massive particles.  
Moreover, $U$ is monotonically increasing at large distances and $U\to -1$ at $r\to \infty$, the stable bound orbits exist for massive particles  even at infinity.

\subsection{Five-center solutions}

For the five  centers, the $z$-axis of ${\mathbb E}^3$ in the Gibbons-Hawking space consists of the six intervals,  
$I_-=\{(x,y,z)|x=y=0,  z<z_1\}$, $I_i=\{(x,y,z)|x=y=0,z_i<z<z_{i+1}\}\ (i=1,...,4)$ and $I_+=\{(x,y,z)|x=y=0,z>z_5\}$. 
As was previously discussed in~\cite{Tomizawa:2019egx},  only particles with the angular momentum of $l_{\phi_2}=0$ can move on $I_1$, $I_3$ , $I_+$ but cannot move on $I_-$, $I_2$ and $I_4$. 
Similarly, only particles with the angular momenta of $l_{\phi_1}=0$ are allowed to move on $I_-$, $I_2$ and $I_4$ but cannot move on $I_1$, $I_3$ and $I_+$.  Here, as typical examples, we study two cases of $j^2\simeq 0.919$ and $j^2\simeq 14.8$ for particles with $l_{\phi_2}=0$.

\subsubsection{$j^2\simeq 0.919$}

\begin{figure}[H]
 \begin{tabular}{cccc}

 \begin{minipage}[t]{0.26\hsize}
\includegraphics[width=4cm,height=4.5cm]{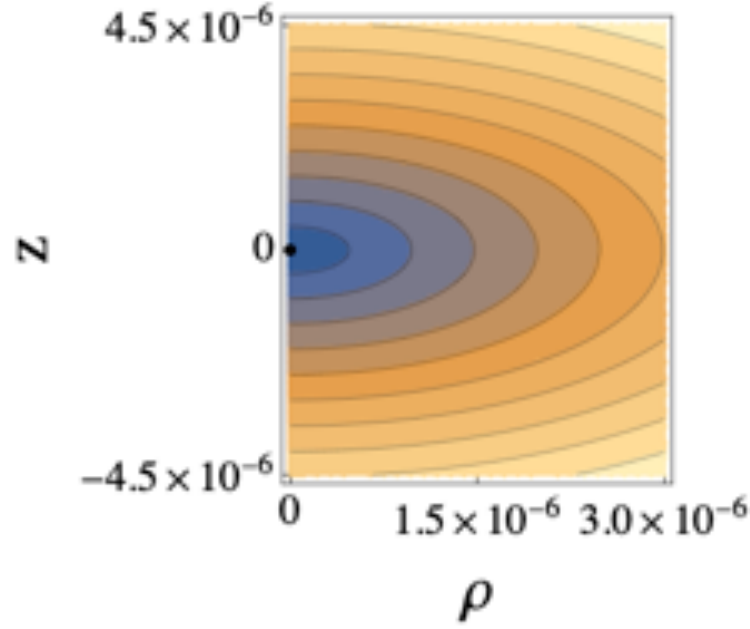}
 \end{minipage} &

 \begin{minipage}[t]{0.26\hsize}
\includegraphics[width=4cm,height=4.5cm]{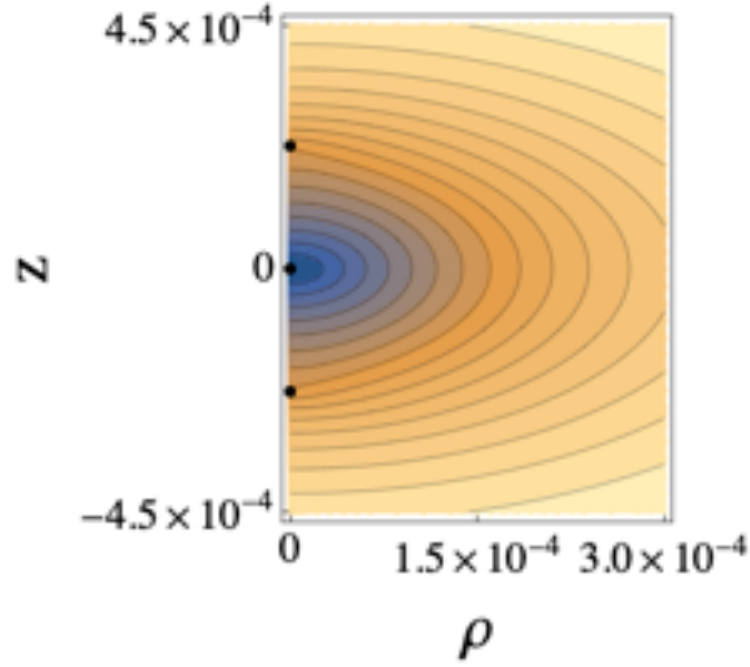}
 \end{minipage} & 
 
 \begin{minipage}[t]{0.26\hsize}
\includegraphics[width=3.7cm,height=4.5cm]{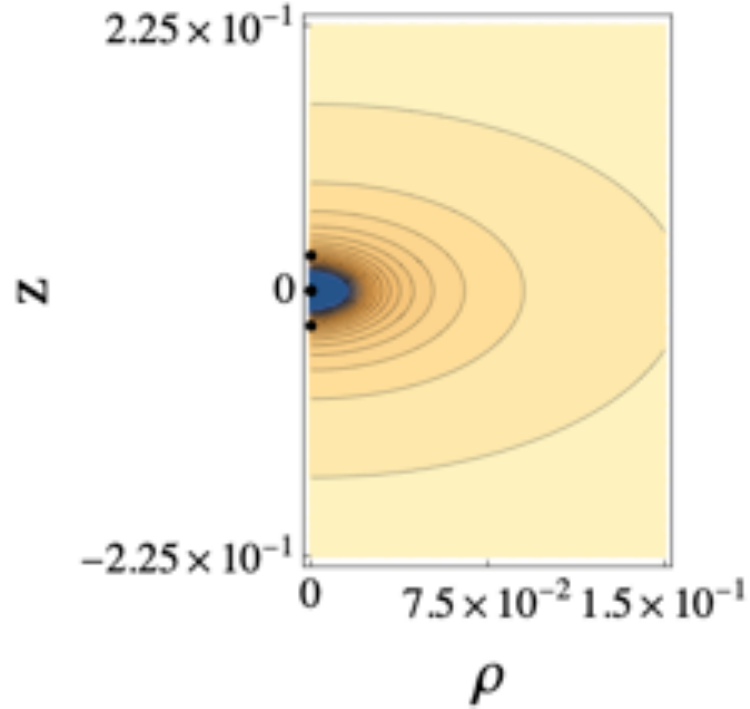}
 \end{minipage} &
 
 \begin{minipage}[t]{0.26\hsize}
 \includegraphics[width=3.7cm,height=4.5cm]{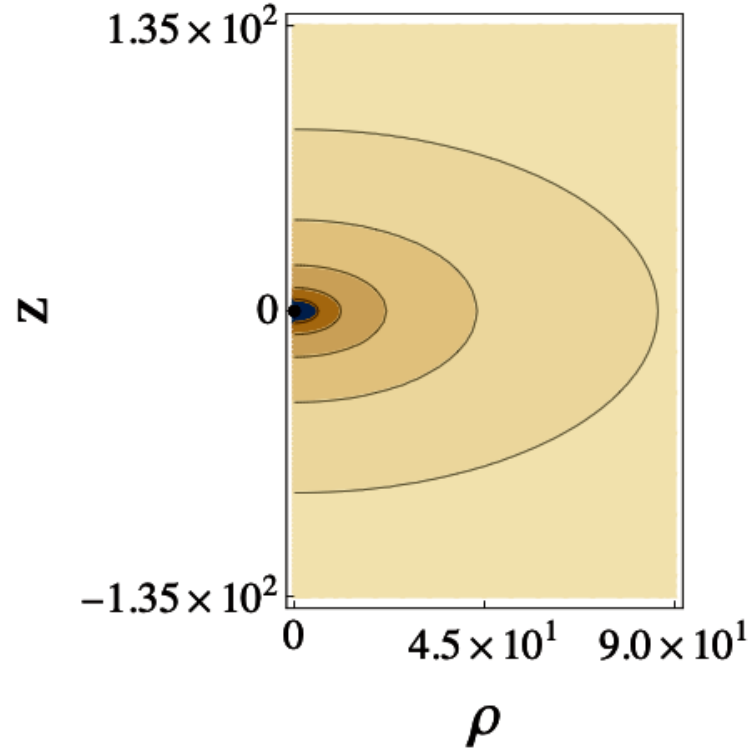}
  \end{minipage}

 \end{tabular}
\caption{The figures show the contours of the potential $U$ under the parameter setting $(k_1,k_2,k_3,k_4,k_5)=(1,-0.192,0,-0.192,1)$ and angular momenta $(l_{\phi_1},l_{\phi_2})=(0,0) $, where $j^2\simeq 0.919$ and  $(a,b)\simeq (0.0296, 0.000227)$.
 The solid curves denote the contours $U={\rm constant}\ (<0)$, and the black circles correspond to the centers at $(\rho,z)=(0,z_i)\ (i=1,\ldots,5)$. 
  }
\label{fig:l10l20_j2=0.919}
\end{figure}

  Figure~\ref{fig:l10l20_j2=0.919} displays the contour plots of $U$ for particles with zero angular momenta, $(l_{\phi_1},l_{\phi_2})=(0,0)$  under the parameter-setting $(k_1,k_2,k_3,k_4,k_5)=(1,-0.192,0,-0.192,1)$, which corresponds to the solutions with $j^2\simeq 0.919$ and  $(a,b)\simeq (0.0296, 0.000227)$, where 
it should be noted that the four figures differ only in the scales of the horizontal axis and vertical axis. 
It can be shown from these figures that  for $(l_{\phi_1},l_{\phi_2})=(0,0)$, $U$ is negative (hence $U$ does not have a $U=0$ curve) and has a local minimum at the center $(0,0)$.  
Therefore,  stable bound orbits do not exist for massless particles but exist for massive particles.  
In particular, massive particles at $(0,0)$ remain at rest.  
Moreover, $U$ is monotonically increasing at large distances and $U\to -1$ at $r\to \infty$, stable bound orbits exist for massive particles  even at infinity.

\begin{figure}[H]
 \begin{tabular}{ccc}
 
 \begin{minipage}[t]{0.3\hsize}
\includegraphics[width=5cm,height=5cm]{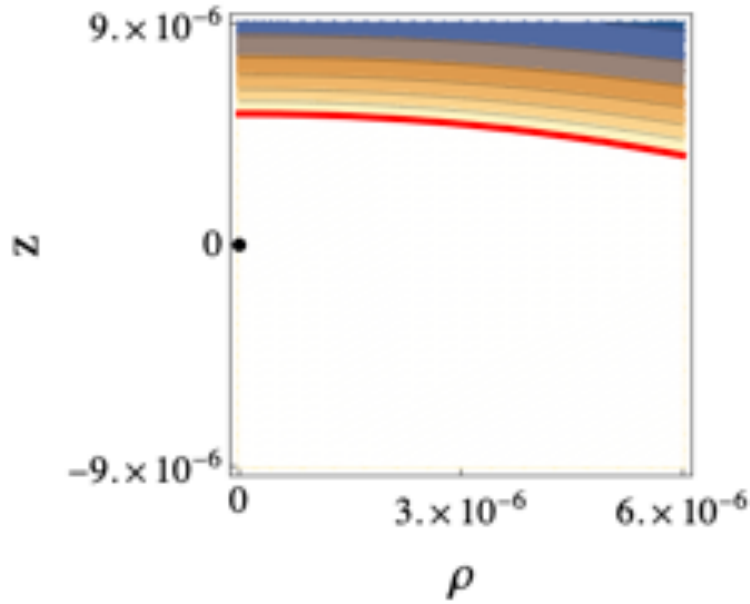}
 \end{minipage} & 
 
 \begin{minipage}[t]{0.3\hsize}
\includegraphics[width=5cm,height=5cm]{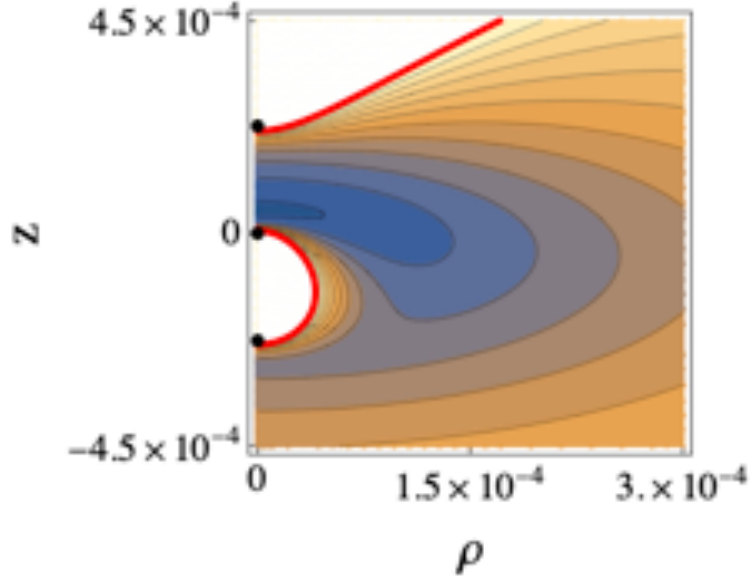}
 \end{minipage} &
 
 \begin{minipage}[t]{0.3\hsize}
 \includegraphics[width=5cm,height=5cm]{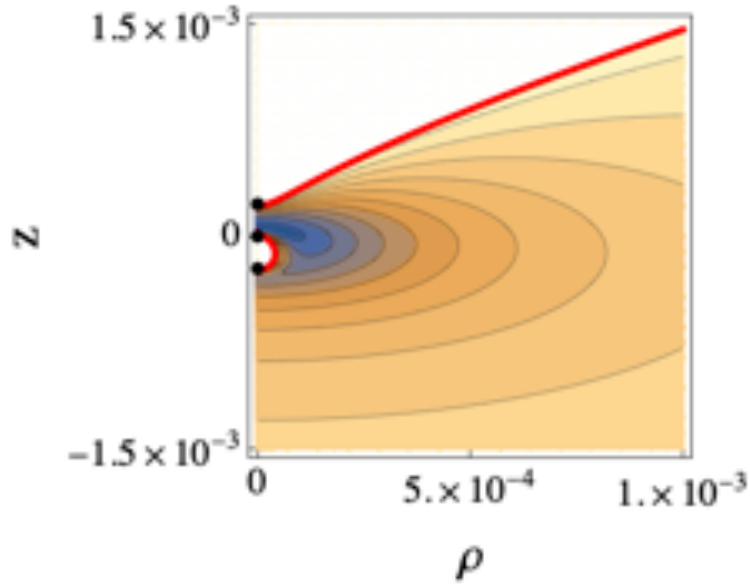}
  \end{minipage} \\
 
 \begin{minipage}[t]{0.3\hsize}
\includegraphics[width=5cm,height=5cm]{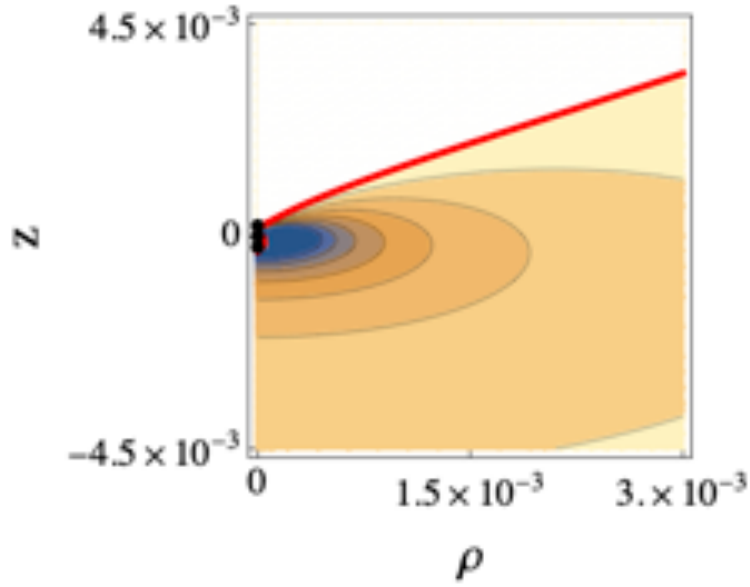}
 \end{minipage} & 
 
 \begin{minipage}[t]{0.3\hsize}
\includegraphics[width=5cm,height=5cm]{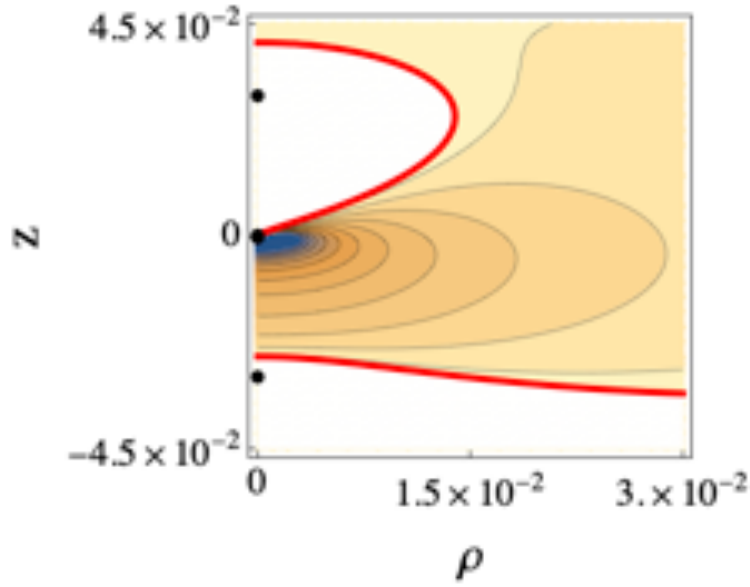}
 \end{minipage} &
 
 \begin{minipage}[t]{0.3\hsize}
 \includegraphics[width=5cm,height=5cm]{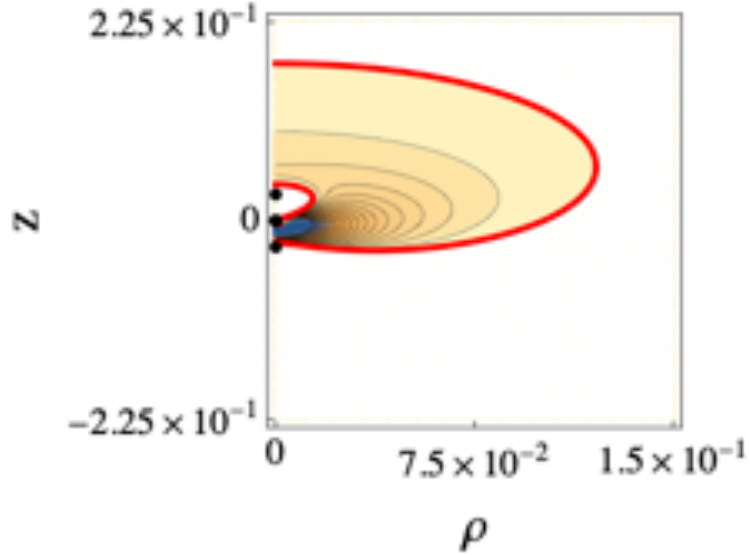}
  \end{minipage} \\
 
  \begin{minipage}[t]{0.3\hsize}
\includegraphics[width=5cm,height=5cm]{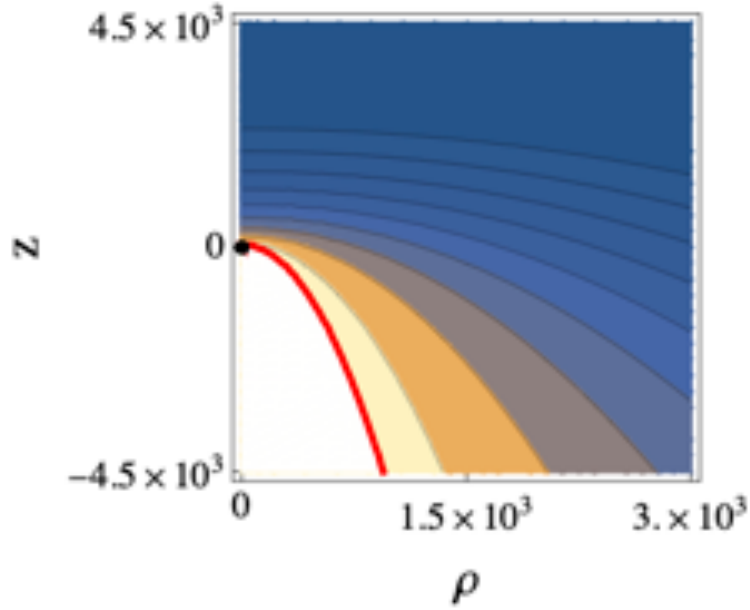}
 \end{minipage} & 
 
 \begin{minipage}[t]{0.3\hsize}
\includegraphics[width=5cm,height=5cm]{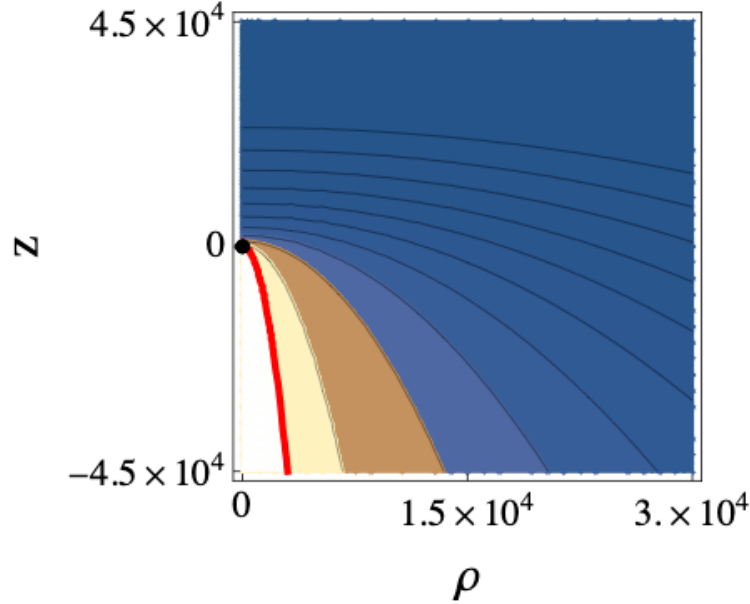}
 \end{minipage} &
 
 \begin{minipage}[t]{0.3\hsize}
 \includegraphics[width=5cm,height=5cm]{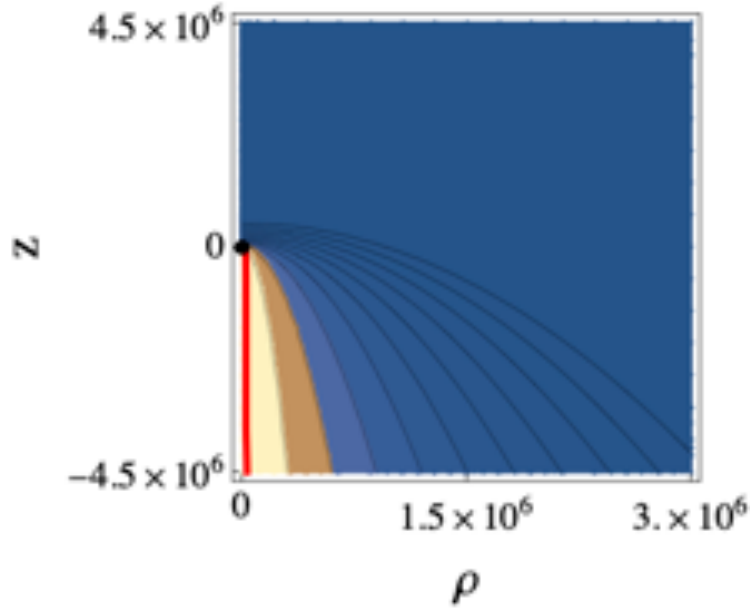}
 \end{minipage} 
  
 \end{tabular}
\caption{The figures show the contours of the potential $U$ under the parameter setting $(k_1,k_2,k_3,k_4,k_5)=(1,-0.192,0,-0.192,1)$ and angular momenta $(l_{\phi_1},l_{\phi_2})\simeq (-14,0) $, where $j^2\simeq 0.919$ and  $(a,b)\simeq (0.0296, 0.000227)$.
The bold solid curves denote $U=0$ and 
 the white regions denote the forbidden regions of $U>0$ where massive and massless particles cannot move. 
 The black circles correspond to the centers at $(\rho,z)=(0,z_i)\ (i=1,\ldots,5)$.  }
\label{fig:l1l20_j2=0.919a}
\end{figure}

Next, we consider the case of $l_{\phi_1}\not=0,l_{\phi_2}=0$ under the same parameter-setting as the above case.
Figure~\ref{fig:l1l20_j2=0.919a} displays the typical contour plots of $U$ for particles with non-zero angular momenta, where we plot the case  $(l_{\phi_1},l_{\phi_2})=(-14,0)$ as an example. 
It should be noted here that each figure differs in only the scales of the horizontal axis and vertical axis.
 For particles with $(l_{\phi_1},l_{\phi_2})=( -14,0)$, there are four $U=0$ curves, 
 (i) the inner $U=0$ curve which surrounds the two centers at $z=z_2\ (\simeq -0.000227)$ and $z=z_3\ (=0)$ and  intersects with $I_1$ and $I_3$
 [the $U=0$ curve in the upper left figure and the lower $U=0$ curves in the upper middle and upper right figures],
 (ii) the inner $U=0$ curve which surrounds the two centers at $z=z_4\ (\simeq 0.000227)$ and $z=z_5\ (\simeq 0.0298)$ and  intersects with $I_3$ and $I_+$
 [the upper $U=0$ curves in the upper middle and the upper right figures, 
 the $U=0$ curve in the middle left figure,
 the upper $U=0$ curve in the middle figure, and 
 the smaller $U=0$ curve in the middle right figure],
 (iii) the intermediate $U=0$ curve which surrounds the two inner $U=0$ curves with the $z$-axis, and  intersects with $I_1$ and $I_+$
 [the larger $U=0$ curve in the middle right figure],  
 (iv) the outer $U=0$ curve which surrounds the intermediate $U=0$ curve and intersects with only $I_+$  
 [the $U=0$ curves in the three lower figures],
  \medskip
 It can be seen from these figures that $U>0\ (<0)$ inside (a little outside) the inner $U=0$ curves, 
 $U<0\ (>0)$ a little inside (a little outside) the intermediate  $U=0$ curve, 
 and 
 $U>0\ (<0)$ a little inside (outside) the outer $U=0$ curve. 
  Thus, $U$  has two negative local minima in the closed region surrounded with the two inner  $U=0$ curves, the intermediate $U=0$ curve and the $z$-axis.
 Therefore,  there are stable bound orbits  for both massive and massless particles in the  region. 
Moreover, a stable circular orbit exists for massive particles at the local minimum of $U$, which is on $I_3$ and hence particles at the point must move along $\partial/\partial\phi_1$.

\subsubsection{$j^2\simeq 14.8$}

\begin{figure}[H]
 \begin{tabular}{ccc}
 
 \begin{minipage}[t]{0.3\hsize}
\includegraphics[width=5cm,height=5cm]{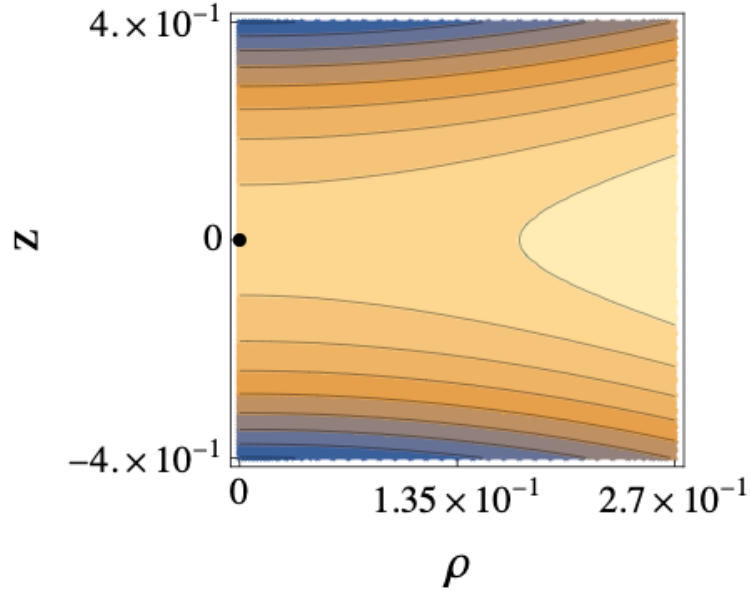}
 \end{minipage} & 
 
 \begin{minipage}[t]{0.3\hsize}
\includegraphics[width=5cm,height=5cm]{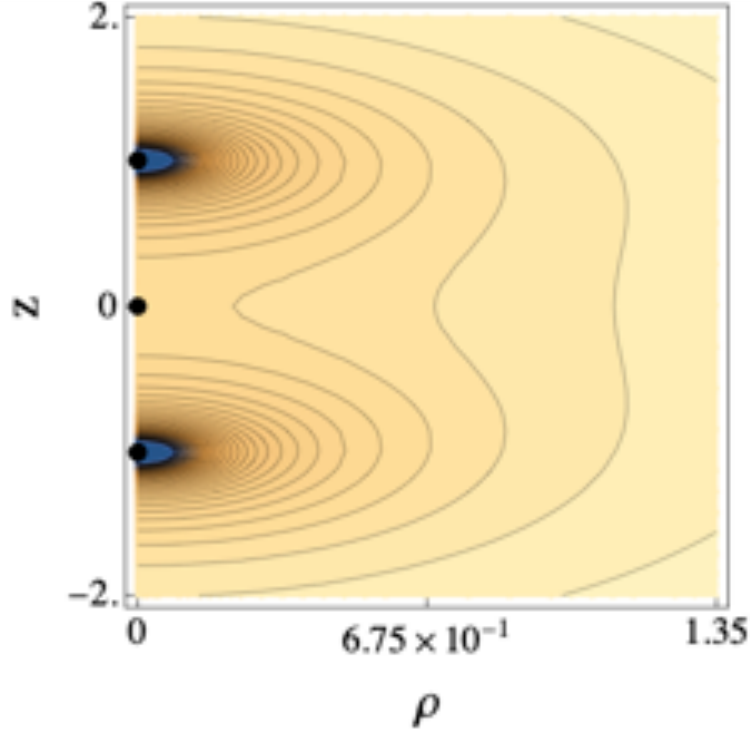}
 \end{minipage} &
 
 \begin{minipage}[t]{0.3\hsize}
 \includegraphics[width=5cm,height=5cm]{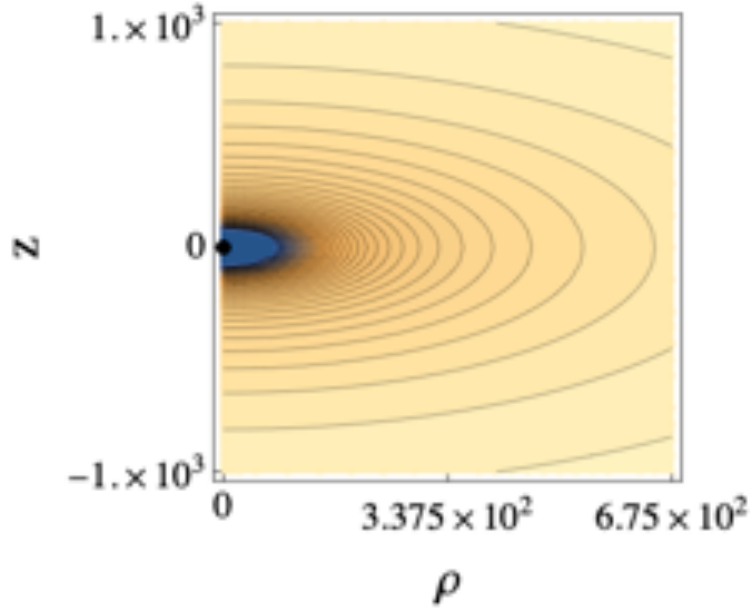}
 \end{minipage}  
 \end{tabular}
\caption{The figures show the contours of the potential $U$ under the parameter setting $(k_1,k_2,k_3,k_4,k_5)=(1,-1.01,0,-1.01,1)$ and angular momenta $(l_{\phi_1},l_{\phi_2})\simeq (0,0) $, where $j^2 \simeq 14.8$ and  $(a,b)\simeq (4.83\times 10^{-7},1.01)$.}
\label{fig:l114a}
\end{figure}

Finally, we study the case $j^2\simeq 14.8$ as the much larger example than the upper bounds for the angular momenta of the BMPV black holes. 
Figure~\ref{fig:l114a} shows the contours of $U$ for $(l_{\phi_1},l_{\phi_2})=(0,0)$ and the parameters $(k_1,k_2,k_3,k_4,k_5)=(1,-1.01,0,-1.01,1)$, which corresponds to the solutions with  $j^2\simeq 14.8$ and  $(a,b)\simeq (4.83\times 10^{-7},1.01)$, where the figures differ only in the scales of the horizontal axis and the vertical axis.
As can be seen from these figures, $U$ does not have a $U=0$ curve, and hence stable bound orbits do not exist for massless particles. 
On the other hand, it can be seen from the middle figure that $U$ has two negative local minima at the $z$-axis,  
so that stable bound orbits  exist for massive particles.

\begin{figure}[H]
 \begin{tabular}{ccc}
 
 \begin{minipage}[t]{0.3\hsize}
\includegraphics[width=5cm,height=5cm]{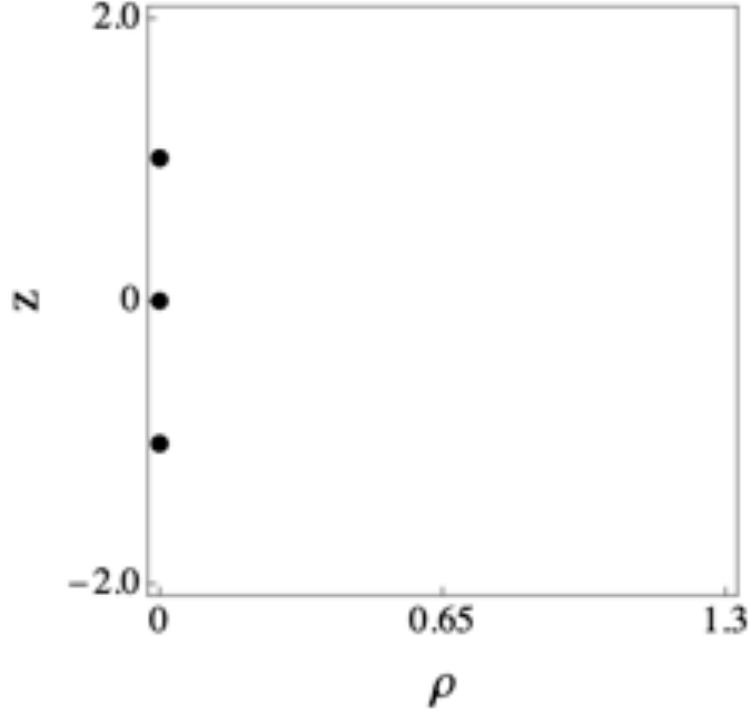}
 \end{minipage} & 
 
 \begin{minipage}[t]{0.3\hsize}
\includegraphics[width=5cm,height=5cm]{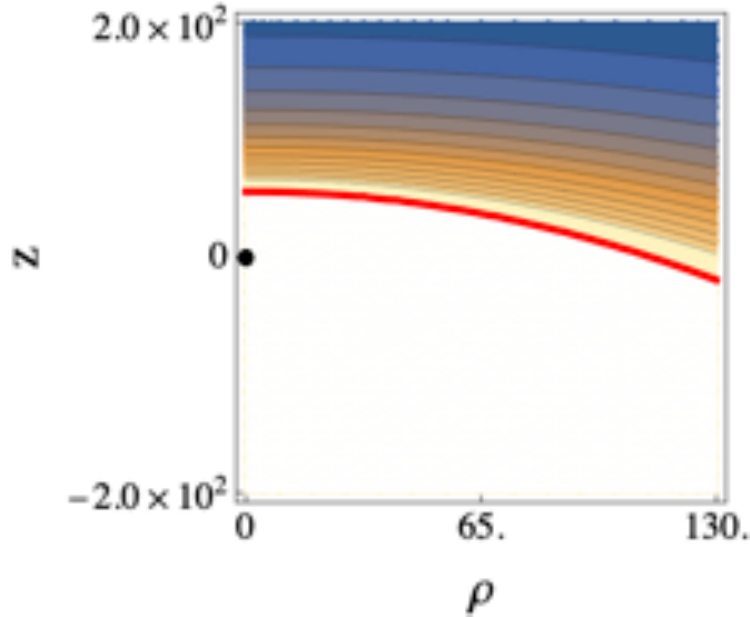}
 \end{minipage} &
 
 \begin{minipage}[t]{0.3\hsize}
 \includegraphics[width=5cm,height=5cm]{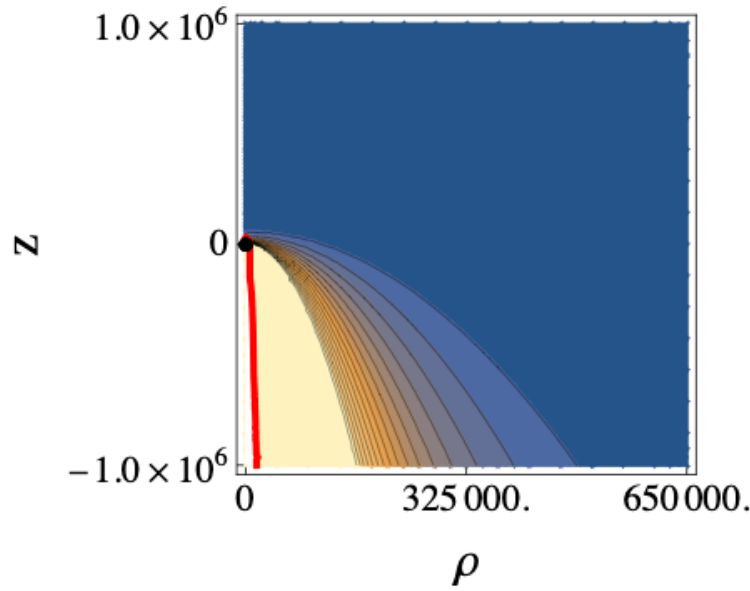}
 \end{minipage}  
 \end{tabular}
\caption{The figures show the contours of the potential $U$ under the parameter setting $(k_1,k_2,k_3,k_4,k_5)=(1,-1.01,0,-1.01,1)$ and angular momenta $(l_{\phi_1},l_{\phi_2})\simeq (-15,0) $, where $j^2 \simeq 14.8$ and  $(a,b)\simeq (4.83\times 10^{-7},1.01)$.}
\label{fig:l114b}
\end{figure}

Figure~\ref{fig:l114b} shows the contours of $U$ for $(l_{\phi_1},l_{\phi_2})=(-15,0)$ under
the same parameter setting as the above. 
As can be seen from these figures, there exists the single non-closed $U=0$ curve which intersects with $I_+$ and extend to infinity.
 $U>0\ (<0)$ inside (outside) the $U=0$ curve, and $U$ does not has a local minimum, i.e.,  bound orbits do not exist for massive/massless particles.    
 Moreover, particles coming in from infinity cannot enter inside the $U=0$ curve, which also occurs for the overrotating BMPV black holes (so-called repulsons).

\section{Summary and Discussions}\label{sec:summary}
We have investigated the existence of stable bound orbits for the massive and massless particles moving in the simplest microstate geometry backgrounds with reflection symmetry (three-center solutions and five-center solutions) in the bosonic sector of the five-dimensional minimal supergravity. 
 In our analysis, reducing the motion of particles to a two-dimensional potential problem, we have plotted the 
 the contours of the potential. 
 More specially, we have shown the following points.
 
 \medskip
 (i) We have numerically shown that the three-center microstate geometries, which must have larger angular momenta than the BMPV black holes,  admit the existence of stable bound for massive particles near the three centers orbits but also even at  infinity. This is quite different from the geodesic behaviors in the five-dimensional  black hole backgrounds. 
We could not confirm the existence of stable bound orbits for massless particles.  
We have found numerically that there is such a finite region near the centers that massive particles with non-zero angular momenta coming in  from infinity cannot reach (the closed white regions in the upper left and upper right figures of Fig.~\ref{fig:n3U}). 
This resembles the repulson behavior of the BMPV black holes with overrotation ($j^2>1$)~\cite{Gibbons:1999uv,Diemer:2013fza}, in which case since the horizon area becomes imaginary, hence ill-defined,  it simply becomes a smooth timelike hypersurface (called a pseudo-horizon). Though the geodesics are complete in such a spacetime, surprisingly, massive and massless particles cannot enter inside the pseudo-horizon.

\medskip
(ii) We have compared the five-center solutions and the BMPV black holes with the same mass and two angular momenta.
In the underrotating BMPV black hole background ($0<j^2<1$), 
stable bound orbits  do not exist outside the horizon for massive and massless particles, 
whereas in the microstate geometries, they exist (near the five Gibbons-Hawking centers) for both massive and massless particle. 
In addition, the solutions also admit the repulson behavior of the overrotating BMPV black holes since 
particles with non-zero angular momenta cannot enter inside the two $U=0$ curves\ (see the closed white regions in the upper middle and middle figures of Fig.~\ref{fig:l1l20_j2=0.919a}). 
Moreover, we have investigated the five-center microstate geometries (with  reflectional symmetry) having angular momenta larger than the BMPV black holes. 
At least, numerically, we could  confirm the existence of stable bound orbits for massive particles with zero angular momenta (not only near the centers but also at infinity) but could not  for massless particles.  
Moreover,  particles with non-zero angular momenta cannot reach near the Gibbons-Hawking centers\ (see Fig.~\ref{fig:l114b}).

\medskip
 It is an interesting issue to analyze more general microstate solutions with a larger number of centers or without reflectional symmetry of centers. These deserve our future works.




\acknowledgments
We thank Takahisa Igata for useful discussion and comments. 
RS was supported by JSPS KAKENHI Grant Number~JP18K13541.
ST was supported by JSPS KAKENHI Grant Number~17K05452 and 21K03560.




\end{document}